\documentclass{article}
\title{Charge conservation and time-varying speed of light}

\author{S. Landau\thanks{Fellow of FOMEC} \emph{F. C. Astron{\'o}micas
y Geof{\'\i}sicas, U. N. de La Plata}\\ 
P. D. Sisterna \emph{Facultad
de C.  Exactas, U. N. de Mar del Plata}\\ and
H. Vucetich\thanks{Member of CONICET} \emph{F. C. Astron{\'o}micas y
Geof{\'\i}sicas, U. N. de La Plata} }

\begin{document}
%\draft
\maketitle

\begin{abstract}
	It has been recently claimed that cosmologies with time
	dependent speed of light might solve some of the problems of
	the standard cosmological scenario, as well as inflationary
	scenarios. In this letter we show that most of these models,
	when analyzed in a consistent way, lead to large violations of
	charge conservation. Thus, they are severly constrained by
	experiment, including those where $c$ is a power of the scale
	factor and those whose source term is the trace of the
	energy-momentum tensor. In addition, early Universe scenarios
	with a sudden change of $c$ related to baryogenesis are
	discarded.
\end{abstract}

%\pacs{23.23.+x, 56.65.Dy}

Since one of the key hypothesis of special relativity is the frame
independence of the velocity of light $c$, it is implied in this
statement the time and space independence of this velocity. As well
established that it may seem, this constancy principle has been
recently contested \cite{Moffat,AlbMag99} to provide an alternative
account of the horizon, flatness and cosmological constant problems
present in the standard big bang scenario. Instead of the superluminal
expansion of the Universe present in inflationary scenarios, a period
in which light traveled much faster than today would explain the
homogeneity we see today in the Universe. Some cosmological models
have also been analyzed afterwards \cite{Barrow99,BranMag99} to test
the dynamical viability of this scenario.

	These ideas are highly provocative, not only from the
observational viewpoint but also from the conceptual one. Indeed one
of the key aspects of Einstein equivalence principle is the
time-independence of the so called ``fundamental constants'' of
physics \cite{Will93}.  The replacement of these parameters by one or
more dynamical fields can lead to time- as well as space-dependent
local fundamental constants. Unification schemes such as superstring
theories \cite{Strings} and Kaluza-Klein theories \cite{KKTheor} have
cosmological solutions in which the low-energy fundamental constants
are functions of time. Usually low-energy phenomena are used to
analize the variation rate of the fundamental constants
\cite{Ellis89}-\cite{Webb00}. 
%Moreover, non-null results have been
%announced  \cite{Webb99} and confirmed while we were correcting this paper \cite{Webb00}.  
If the cosmological dynamics of a
field is such that its large-scale value is invariant under local
Lorentz transformations, or if the local coupling with matter is
strong enough so that it depends on the local environment (\emph{e.g.}
electromagnetism with the absorber condition), then the local field
equations will be Lorentz invariant. If on the other hand the
cosmological evolution is non-trivial and the field couples softly
with the local matter, it will act as an external bath, breaking local
Lorentz invariance. A variable speed of light theory may belong to the
latter set of theories. Any VSL theory poses an additional problem,
namely that $c$ is a dimensional constant, and talking about a varying
dimensional constant is not an invariant statement: we can change our
units and obtain a different time dependence of such a parameter. Of
course, once we fix our units, every claim about a dimensional
parameter is an invariant claim, since we are implicitly referring to
a dimensionless ratio: that between the parameter and the unit
\cite{TVOFC1,TVOFC2}.

	Any scientific theory has to be stated in clear and precise
terms. Beckenstein's theory of a variable fine structure constant was
based on Lorentz invariance, explicitly protecting charge
conservation. In the case of VSL theories, local Lorentz invariance is
relaxed, and  the inhomogeneous Maxwell equations are assumed to be
\cite{AlbMag99}
\begin{equation}
\frac{1}{c} \partial_\mu\left(cF^{\mu\nu}\right) = 4\pi j^\nu,
					\label{ModMaxwell}
\end{equation}
where $j^\mu = (\rho, {\bf j}/c)$ is the electric charge current. In
reference \cite{AlbMag99} it was suggested that charge is conserved,
implying a variation of the fine structure constant $\alpha =
\frac{e^2}{\hbar c}$. The constancy of $e$ can be derived, for
instance, from Dirac's equation, written in Hamiltonian form:
\[
i\hbar \partial_t \psi = -(i\hbar\mbox{\boldmath{$\alpha$}} \cdot
\mbox{\boldmath{$\nabla$}} + {\rm h.c.}) \psi + mc^2 \psi 
\]
which implies that $Q = e\int_V \psi^*\psi d^3x$ is conserved. The
above form is, however, not unique since powers of $c$ can be
introduced in the equation in several ways, followed by appropriate
symmetrization. 

	However, from equation (\ref{ModMaxwell}) it can be easily seen that
$j^\mu$ is no longer a conserved current, but satisfies the equation:
\begin{equation}
4\pi \partial_\mu (cj^\mu) =  \partial_\mu \partial_\nu(cF^{\mu\nu})
					\label{NonConsCharge}
\end{equation}

	The right hand side is not null, because the partial
derivatives do not commute:
\[
[\partial_0, \partial_i] = \frac{c_{,i}}{c^2} \partial_t
\]
It is usually assumed that $c$ is a function of a scalar field and
that it depends only on time in the comoving cosmological frame. In
the local frame of the solar system, moving with a velocity $v$ with
respect to the cosmological frame, a small space dependence will
arise, with gradients $O(v/c)$ with respect to the time derivative,
which can be neglected for the present purposes. So, the right hand
side of equation \ref{NonConsCharge} is effectively zero.  The left
hand side, however, is not a four divergence, because $\partial_{x^0}
= 1/c(t) \partial_t$. The fully expanded expression is:
\begin{equation}
\partial_t \rho + \frac{\dot{c}}{c}\rho + \nabla\cdot{\bf j} = 0
				\label{DifConsCharge}
\end{equation}

	If equation (\ref{DifConsCharge}) is integrated over a volume
$V$ containing the charges, we obtain
\begin{equation}
\frac{\dot{Q}}{Q} = - \frac{\dot{c}}{c}
				\label{VarCharge}
\end{equation}
where $Q = \int_V \rho d^3x$ is the total electric charge. This is our
main result.

	Equation (\ref{VarCharge}) provides  very stringent tests of
the variation of $c$, since there have been many experiments to test
the conservation of charge \cite{Okun92}. Depending on the details of
the theory, several cases arise.

	If we assume, as it is usually done in this context
\cite{AlbMag99,Barrow99} that the electron charge $e$ is constant,
charge conservation can only be broken by processes that change charge
discontinuously, such as the dissapearance of electrons or the
transformation of neutrons into protons. A generic model for these
processes has been proposed in references \cite{Magueijo00,CQG16},
under the assumption of total energy conservation.  The first three
entries of Table \ref{NCQTab} show some sample upper limits obtained
from these processes with different techniques and hypothesis. These
upper limits on $\dot{c}/c$ are much smaller than those obtained by a
direct measurement, for instance in reference \cite{Prestage95},
namely $\mid
\dot{c}/c \mid < 10^{-13}\; {\rm yr}^{-1}$.

	On the other hand, if $e$ varies continuously in such a way
that $ce$ is conserved, then $\alpha \propto c^{-3}$ and strict limits
can be obtained from geophysical or astronomical data, such as the
Oklo phenomenon \cite{Shlyakhter76,DamDys96} or the line spectra of
distant quasars \cite{CyS,Pothekin98}. Furthermore, evidence for the time variation of the fine structure constant has been claimed by Webb et al. \cite{Webb99} and confirmed while we were correcting this paper \cite{Webb00}. Thus, if the systematic errors are well estimated, the requirement of conservation of charge suggests that, the mechanism for $\alpha$ variation should not be driven by the change in the speed of light.

	These limits discard a great number of cosmological models
with varying velocity of light. For instance, the family introduced in 
\cite{Barrow99} parameterizes light velocity in the form:
$					%\begin{equation}
c = c_0 \left(\frac{a}{a_0}\right)^n	%\label{BarrowMod}
$					%\end{equation}
with $a$ the cosmological scale factor. Even though, \cite{Barrow99} was not proposing that this behaviour exists at all times, \cite{BarrMag} suggests that this solutions could be extended to radiation and matter-dominated universes. It was shown in reference
\cite{Barrow99} that $n < -1/2$ is necessary to solve the flatness and
horizon problems, and $n < -3/2$ solves the cosmological constant
problem.  
%However, in these models the following equation holds:
%\begin{equation}
%\frac{\dot{c}}{c} = n \frac{\dot{a}}{a} = n H_0
%					\label{CondOn-n}
%\end{equation}

On the other hand, any complete and consistent VSL
theory will predict dynamically the value of c via a wave-like equation for
$\psi =\ln \frac{c}{c_0}$, whose source term is
proportional to the trace of the energy-momentum tensor $T$. Thus, in the
neighborhood of a quasi-static system, such as a star or a virialized
galaxy cluster, a generic expresion for a varying $c$ will be \cite{Magueijo002}:
\begin{equation}
c =c_c(t)\left(1-\frac{\lambda G M}{c_0^2 r}\right) 
\end{equation}

Here $\lambda$ is a constant that depends on the specific VSL theory.
Furthermore, in the limit $r \rightarrow \infty$ , the expression for $c$ reduces to the cosmological one ($c \rightarrow c_c(t)$).

For the purposes of this paper, our interest is focused on violation of charge conservation. Expanding equation(\ref{NonConsCharge}) we obtain:

\begin{equation}
4\pi\partial_\mu j^\mu = -4 \pi \frac{\partial_\mu c}{c} j^\mu -
\partial_t c \frac{\vec\nabla c . \vec E}{c^3}+
\frac{\vec \nabla . c}{c} \partial_0\vec E
\end{equation}

In a static situation $\vec j = 0$ and $\partial_0 E = 0$, thus the
effects of $\vec \nabla c $ are of second order and can be
neglected. Hence, there are two contributions to the violation of
charge conservation, one accounts for time-variations over
cosmological time-scales and the other for the motion of the Earth
with respect to massive bodies such as clusters or galaxies:
\begin{equation}
\frac{\dot c}{c} = \left(\frac{\dot c}{c}\right)_{cosmological}
+\left(\frac{\dot c}{c}\right)_{local} = n \frac{\dot a}{a} +
\frac{\lambda G M}{c_0^2 r^3} \vec r . \vec v \label{cvar}
\end{equation}
where $\vec v$ is the velocity of the Earth with respect to the
lump. ($\vec v \sim 1000 \frac{km}{seg}$ for the Virgo supercluster).

From the first term of equation (\ref{cvar}) we find the limits of table
\ref{NCQTab} on $n$, which contradict the above requirements. (We use
$H_0 = 65\; {\rm km/s/Mpc} = 6.65 \times 10^{-11}\; {\rm
yr}^{-1}$). Similar bounds can be obtained for other similar models, such
as those studied in reference \cite{BarrMag} (See also \cite{Chimento}).

	For the Virgo supercluster, $\frac{G M}{c^2_0 r} \sim 2 \times
10^{-7}$.  The last three results in table \ref{NCQTab} for $\lambda$
are obtained from the second term of equation (\ref{cvar}) . These
results rule out essentially any VSL whose source of variation is the
trace of the energy-momentum tensor $T$.

	Finally, models similar to the original Albrecht-Magueijo one
\cite{AlbMag99}, involving a sudden change of $c$ between two
different constant values in the very early Universe, are not affected
by the above limits. Orito and Yoshimura \cite{Orito85} observed that
if charge conservation is broken in the very early Universe, a large
charge excess should have been formed through a mechanism similar to
that of baryogenesis: violation of $Q$, $C$ and $CP$ conservation
while the system is out of thermodynamic equilibrium
\cite{KT83,KT90}. In the above mentioned models, the net
charge excess will be produced by way-out-of-equilibrium production and
decay of heavy mesons \cite{KT90}.

	Let $X$ be an unstable meson that produces a mean baryon
number $\epsilon_B$ and a mean charge excess $\epsilon_Q$ per decay,
and assume that matter is created during the charge transition
period. Then, the equations for the evolution of the number densities
of $X, B, Q$ will be \cite{AlbMag99,Barrow99,KT90}:
\begin{eqnarray*}
{(a^3 n_X)}^\bullet + \lambda_X (a^3 n_X) &=& 
	\frac{3Kc\dot{c}}{4\pi G m_X} a\\
\dot{(a^3 n_B)}^\bullet &=& \epsilon_B \lambda_X n_X a^3\\
\dot{(a^3 n_Q)}^\bullet &=& \epsilon_Q \lambda_X n_X a^3
\end{eqnarray*}
In these equations, $K$ is the curvature parameter, $m_X$ is the mass
of the X meson, and all densities are effective densities (particle -
antiparticle). Let us assume that the change in $c$ occurs in a short
interval of time $t_c<<\tau<<\frac1{\lambda_x}$. Charge conservation
will be broken only during this interval, but the $X$ meson decay will
always produce a baryon excess. With these hypothesis, the above
equations have the following solutions:
\begin{eqnarray*}
a^3 n_X &\simeq \frac{3K(c_0^2 - c_P^2)}{8\pi G m_X} a(0)
	e^{-\lambda t} &= a^3(0) n_X^0 e^{-\lambda
	t}\\
a^3 n_B &\simeq \epsilon_B a^3(0) n_X^0 \left(1 -  e^{-\lambda t}\right)
	&\rightarrow \epsilon_B a^3(0) n_X^0\\
a^3 n_Q &\simeq \epsilon_Q \lambda_X \tau a^3(0) n_X^0 &\simeq
	\frac{\epsilon_Q}{\epsilon_B} \lambda_X \tau a^3 n_B
\end{eqnarray*}
After the transition, $n_Q$ will be fixed but $n_B$ will be diluted
from the above estimate by thermal processes \cite{KT90}. So we
finally get a lower bound on the charge excess:
\begin{equation}
\left|\frac{n_Q}{n_B}\right| \sim \frac{\epsilon_Q}{\epsilon_B}
	{\lambda_X\tau} \label{ChargeExcess}
\end{equation}
	
	As we have mentioned before, we expect on general grounds that
$\tau > t_{Pl}$, while $\epsilon_Q \sim \epsilon_B$, since these
fractions depend both on the $C$ and $CP$ breaking terms in the
lagrangean. Thus, equation (\ref{ChargeExcess}) predicts a firm lower
limit for the charge excess.  Orito and Yoshimura \cite{Orito85} have
given limits on any charge excess in the Universe, shown in table
\ref{NCQTab}. These limits are many orders of magnitude below the
prediction of equation (\ref{ChargeExcess}). The last column of the
table shows rough estimates of $\tau$ taken from the observational
limits, assuming $1/\lambda_X \sim t_{GUT}$.

%	Finally, it may happen that local physics may not reflect the
%cosmological change of $c$. This is because many VSL theories, when
%consistently developed, assume that $c$ (or some auxiliary field
%$\psi$, from which the light velocity can be obtained in the form $c =
%c_0 f(\psi)$) obey some wave-like equation whose source term is
%proportional to the trace of the energy-momentum tensor $T$. In the
%neighborhood of a quasi-static system, such as a star or a virialized
%galaxy cluster, that equation should reduce to the Poisson equation,
%with an asymptotic behavior for $c$:
%\begin{equation}
%\frac{c(r)}{c_0} \sim \lambda \frac{G M}{c^2_0 r}
					\label{Qloc-Pot}
%\end{equation}
%where $\lambda \sim n$ is related to the coupling constant of $c$ to
%matter. The motion of the Earth with respect to those massive bodies
%with a velocity ${\bf v}$ will introduce an effective variation of $c$
%in the form:
%\begin{equation}
%\frac{\dot{c}}{c_0} = -\lambda  \frac{G M}{c^2_0 r^3} 
%	{\bf r \cdot v}
					\label{Qloc-Var}
%\end{equation}

%	For a nearby galaxy cluster, the velocity is dominated by the
%Hubble flow \cite{Oklo-eq} and we obtain for the charge production:
%\begin{equation}
%\frac{\dot{Q}}{Q} = \lambda \frac{G M}{c^2_0 r} H_0
%					\label{Qloc-Cluster}
%\end{equation}

%	For the Virgo supercluster, $\frac{G M}{c^2_0 r} \sim 2 \times
%10^{-7}$ From equation (\ref{Qloc-Cluster}) we obtain the last three results in
%table \ref{NCQTab} for $\lambda$. These results rule out essentially
%any VSL whose source of variation is the trace of the energy-momentum
%tensor $T$.

	Although these results do not rule out all varying velocity of
light theories, they put very stringent bounds on them through the
conservation of charge requirement.  Moreover, similar bounds will
hold for any theory with varying speed of light velocity in the early
universe. These bounds, which may be lowered through improvements in the
experimental techniques \cite{ALV},  will lead into deeper
understanding of these interesting theories.

        The authors want to thank Diego F. Torres for many interesting
discussions and advice.  The authors acknowledge partial economic
support through the project \texttt{011/G035}, Universidad Nacional de
La Plata.

\begin{table}
\begin{tabular}{lcccc}
Process& Ref.  & Datum & $\mid\dot{Q}/{Q}\mid$ (y$^{-1}$) & Param. \\
\hline
 & & $\tau$ (y)& & $\mid n\mid$ \\ 
\hline 
$^{71}{\rm Ga} \rightarrow ^{71}{\rm Ge}$ & \cite{NBG96} & $\geq 3.5
	\times 10^{26}$ & $\leq 2.9 \times 10^{-27}$ & $< 5\times
	10^{-17}$ \\  
$ e \rightarrow \nu_e \gamma $ & \cite{Avign86} & $\geq 2.4 \times 10^{25}$
	& $ \leq 4.2 \times 10^{-26} $ &$< 7\times 10^{-16}$ \\ 
$ e \rightarrow {\rm any}$& \cite{Reusser91} & $\geq 2.7\times10^{23}$ & $
	\leq 3.7\times10^{-24}$ & $<6\times10^{-14}$ \\ 
\hline 
& & $\mid\Delta\alpha/\alpha\mid$ & & \\ 
\hline 
Oklo phenomenon& \cite{DamDys96} & $\leq 1.2 \times 10^{-7}$ & $ \leq
	2.0 \times 10^{-16} $ &$< 3\times 10^{-6}$ \\
Quasar absorption systems & \cite{CyS} & $\leq 3.5 \times 10^{-4}$ & 
	$ \leq 1.18 \times 10^{-13} $ & $< 2\times 10^{-3}$\\  
Quasar absorption systems & \cite{Pothekin98} & $\leq 6 \times 10^{-5}$ & 
	$ \leq 5\times 10^{-15} $ & $< 7\times 10^{-5}$\\ 

\hline & & $n_Q/n_B$ & & $\tau/t_{Pl}$\\ 
\hline 
Coulomb force smaller than&  & &
	&  \\ 
Newton force in stars &\cite{Orito85} &$ < 10^{-18}$ & ---  & $< 10^{-7}$\\ 
CMB anisotropy & \cite {Orito85} & $ <2\times 10^{-20}$ & --- & 
	$ < 2\times 10^{-9}$ \\ 
Cosmic ray isotropy & \cite{Orito85} &$< 10^{-29}$ & --- & $< 10^{-18}$\\
\hline
 & & $\tau$ (y)& & $\mid \lambda \mid$ \\ 
\hline 
$^{71}{\rm Ga} \rightarrow ^{71}{\rm Ge}$ & \cite{NBG96} & $\geq 3.5
	\times 10^{26}$ & $\leq 2.9 \times 10^{-27}$ & $< 2\times
	10^{-10}$ \\  
$ e \rightarrow \nu_e \gamma $ & \cite{Avign86} & $\geq 2.4 \times 10^{25}$
	& $ \leq 4.2 \times 10^{-26} $ &$< 2\times 10^{-9}$ \\ 
$ e \rightarrow {\rm any}$& \cite{Reusser91} & $\geq 2.7\times10^{23}$ & $
	\leq 3.7\times10^{-24}$ & $<2\times10^{-7}$ \\ 
\end{tabular}
\caption{Upper limits on charge non-conservation. The columns show the
process considered, the corresponding references, the observational
data, the charge non-conservation upper bound and the limits for the
model parameters.}
\label{NCQTab}
\end{table}
\end{document}